\documentclass[11pt,twoside]{article}

\usepackage{asp2010}
\usepackage{graphicx}

\resetcounters

\aspvoltitle{Galaxy Mergers in an Evolving Universe}

\begin{document}

\title{Experiments With IDENTIKIT}
\author{Joshua E. Barnes$^1$ and George C. Privon$^2$
\affil{$^1$Institute For Astronomy, 2680 Woodlawn Drive, Honolulu,
       HI\ \ 96822, USA\\
       $^2$Department of Astronomy, University of Virginia,
       530 McCormick Road, Charlottesville, VA\ \ 22904, USA}}

\begin{abstract}
IDENTIKIT was originally developed as a fast approximate scheme for
modeling the tidal morphology and kinematics of disk galaxy encounters
and mergers.  In this form, it was first used to implement an
interactive modeling tool for galaxy collisions; tests with artificial
data showed that the morphology \textit{and} kinematics of merging
galaxies strongly constrain their initial conditions.  This tool is
now being applied to real galaxies.  More recently, IDENTIKIT has been
used to develop a mapping from the present state of a tidal encounter
back to the initial conditions; this offers a way to partly automate
the search for dynamical models of galaxy encounters.  Finally,
IDENTIKIT's theoretical applications include a comprehensive way to
evaluate the mass and extent of tidal features as functions of halo
structure.
\end{abstract}

\section{Introduction}

The tidal theory of galaxy encounters explains the morphological and
kinematic features of ``peculiar'' galaxies \citep{A66} as consequences
of gravitational interactions between previously normal disk galaxies
\citep[hereafter TT72]{TT72}.  While this theory has strong
theoretical foundations, it gained considerable credibility from
TT72's plausible simulations of four well-known interacting galaxies.
In the decades since TT72, the number of systems with detailed
dynamical models has gradually increased.  There are several reasons,
beyond testing the tidal theory, to create models of specific systems:
such models (1) help interpret complex three-dimensional morphology,
(2) provide access to the time domain, and (3) may be used to test
sub-grid models of star formation, feedback, AGN fueling, etc.  Thus,
a dynamical merger model can provide a framework unifying disparate
observations into a coherent picture.

Finding initial conditions which reproduce a specific pair of
interacting galaxies is a time-consuming business.  The parameter set
is large, and parameters interact in complex ways.  Hours may be
required to run a single self-consistent calculation, and weeks of
trial and error may be needed to obtain an acceptable match.

Sixteen parameters specify an encounter between two galaxies; these
fall into three distinct groups as follows.  (1) The initial orbit is
specified by the pericentric separation $p$, mass ratio $\mu$, and
eccentricity $e$. (2) Disk orientations are specified by inclination
angles $i$ and azimuthal angles $\phi$.  (3) The mapping from a
simulation to observables is specified by the time since pericenter
$t$, three viewing angles $\theta_\alpha$, length and velocity scale
factors $\mathcal{L}$ and $\mathcal{V}$, position zero-point
$\mathbf{r}_{0}$ and velocity zero-point $v_{0}$.  Together, groups
(1) and (2) specify the initial conditions.

The criteria used to decide if a model matches the observations are
subtle.  As a rule, it's not enough to reproduce the tidal morphology
of an interacting pair of galaxies, since different encounter
geometries may produce identical morphologies \citep[e.g.][]{B11}.
Kinematic information on tidal bridges and tails provides much
stronger constraints.  Interferometric H{\small I} data, which traces
both tidal morphology \textit{and} kinematics, is often used to
constrain merger models; H$\alpha$ may also be useful if tidal
features contain emission-line regions.  Molecular and/or stellar
absorption lines can also provide constraints, but mapping extended
structures in these lines is expensive.  Models and observations are
typically compared visually, for example by overplotting particles on
various projections of a H{\small I} data cube \citep[e.g.][]{HM95}.
Since interstellar material converts between molecular, atomic, and
ionized phases by processes outside the scope of purely dynamical
models, visual inspection may be more reliable than quantitative
measures derived by differencing model and observational data cubes.
However, visual inspection is tedious and inherently subjective;
robust quantitative methods would certainly be welcome.

Some progress has been made using genetic algorithms to automate the
job of modeling interacting galaxies \citep[e.g.][]{W98, TK01, S+10}.
Interesting results have been obtained, but most tests to date have
searched only limited subsets of the relevant parameters and made
little use of kinematic information.

\section{IDENTIKIT~1}

IDENTIKIT simulations combine test-particle and self-consistent
techniques \citep{BH09}.  Each galaxy is modeled by an initially
spherical configuration of massive particles with cumulative mass
profile $m(r)$, in which is embedded a spherical swarm of massless
test particles on initially circular orbits.  Two such models with
mass ratio $\mu$ are launched towards each other on an orbit with
eccentricity $e$ and pericentric separation $p$.  During the ensuing
encounter, the massive components interact self-consistently, closely
approximating the time-dependent potential and orbit decay of a fully
self-consistent galactic collision.  The test particles mimic the
tidal response of embedded discs with all possible spin vectors; once
such a simulation has been run, selecting the appropriate subset of
test particles yields a good approximation to the tidal response of
any particular disc.

Using this scheme, trial-and-error modeling of interacting galaxies
becomes much less tedious.  \citet{BH09} constructed an artificial
data set of $36$ parabolic ($e = 1$) encounters between equal-mass
($\mu = 1$) disk galaxies.  The encounter geometries, pericentric
separations, times since pericenter, viewing directions, and scale
factors were chosen at random.  With no knowledge of the actual
parameter values, IDENTIKIT~1 was used to search for models
reproducing the observable morphology and kinematics of these $36$
systems.  Some $30$ cases were successfully reconstructed.  In these
cases, all unknown parameters were well-constrained; for example,
encounter geometry and viewing direction were recovered with median
errors of $< 15^\circ$.

\subsection{Modeling real galaxies}

Having passed ``laboratory'' tests, IDENTIKIT is ready to apply to
real galaxies.  This immediately raises several new issues.  First is
the limitations of the observational data.  For most systems, H{\small
I} is the only available tracer of large-scale morphology \textit{and}
kinematics.  In many cases, the resolution and signal-to-noise of the
data barely suffice to trace tidal structures.  Second, the
observational data may display features outside the scope of simple
test-particle models.  For example, H{\small I} data often includes
absorption features, and it's not clear if or how such features can be
matched by the simulations.  Third, the mass models used in the
IDENTIKIT simulations may not be good approximations to the actual
mass distributions of the galaxies involved in an encounter.  At this
point, the simulations use generic disk galaxy models; tailoring these
models to specific systems is an important challenge.

\begin{figure}[t!]
  \begin{center}
    \includegraphics[width=4.25in]{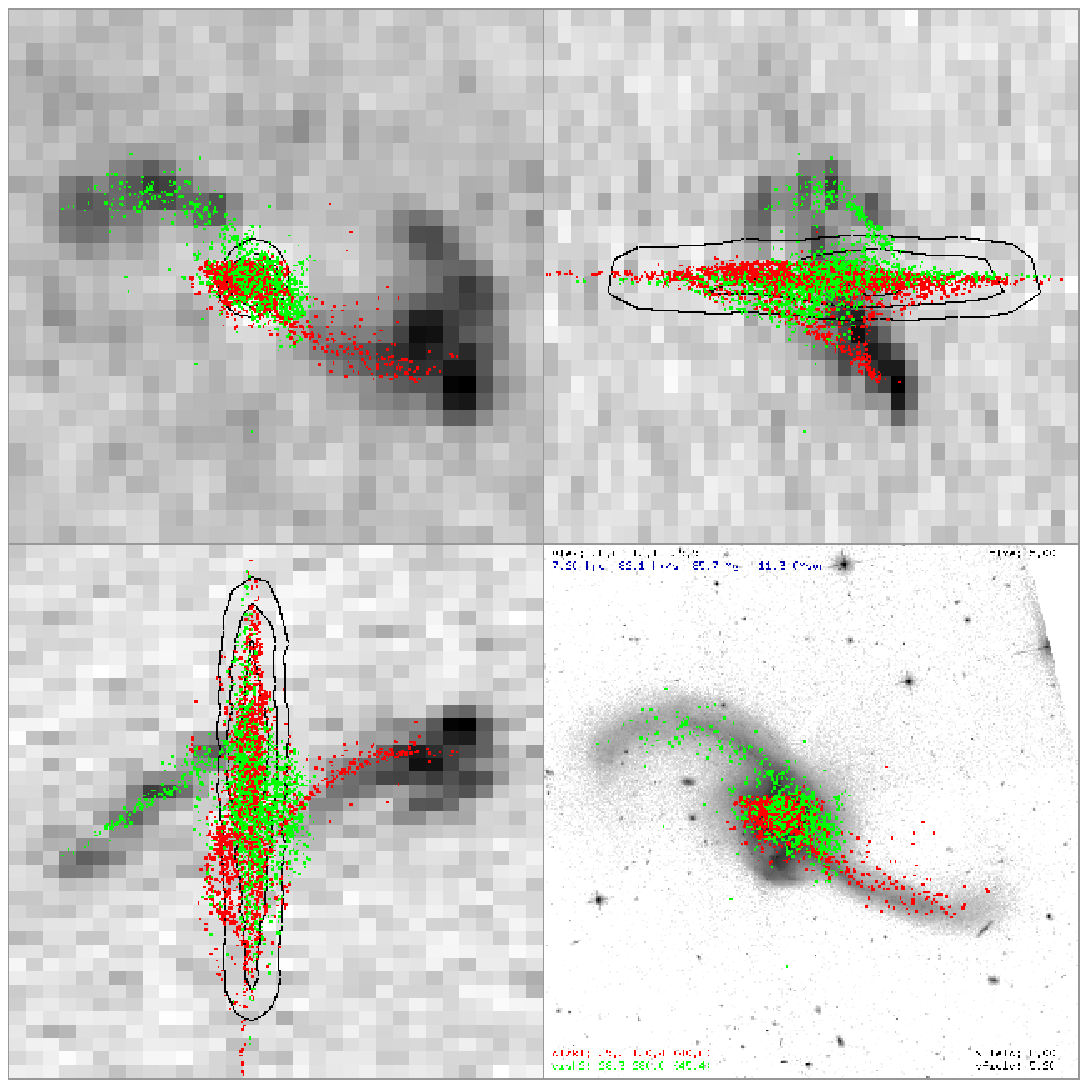}
  \end{center}
  \caption{NGC~2623.  Top-left and bottom-right panels show H{\small
  I} and optical images, respectively; North is up and West is right.
  Top-right and bottom-left panels show H{\small I} position-velocity
  diagrams; velocity increases to left and down, respectively.
  Grey-scale shows emission, while contours show H{\small I}
  absorption.  Points show a preliminary IDENTIKIT~1 model (Privon et
  al., in preparation).}%
  \label{fig1}%
\end{figure}

Fig.~\ref{fig1} presents a preliminary model of NGC~2623
(Privon et al., in preparation) which illustrates all three of these
issues.  First, while H{\small I} is definitely detected in the tails
\citep{HY96}, its emission is easier to map by taking the
\textit{maximum} voxel value along each line of sight through the data
cube; summing the emission tends to produce noisy results.  It's not
entirely clear how to compare the present maps to the the projected
particle distribution, which most naturally corresponds to summed
emission.  In addition, H{\small I} emission is not detected from the
bright star-forming region to the south of the main body of NGC~2623;
optical spectroscopy may help to determine kinematics in this region.
Second, the system exhibits H{\small I} absorption, presumably due to
neutral hydrogen silhouetted against NGC~2623's AGN \citep{E+08}.
It's interesting that the velocity width of this absorption feature is
fairly well reproduced by the width of the particle distribution, but
not entirely obvious that this is a success of the model since
H{\small I} on the far side of the nucleus does not contribute to the
absorption.  Third, while NGC~2623 probably results from a merger of
comparable galaxies, the hook-shaped tail to the West might be better
reproduced if its parent galaxy's disk was initially larger than its
partner's.

Further experience modeling a number of different systems is needed to
explore these issues.  It would also be interesting to combine
H{\small I} with optical data, and to include other velocity tracers
such as H$\alpha$ or CO.  We are currently modeling Arp~240
(NGC~5257/8); other systems on our short-list include NGC~34 and
NGC~3256.  Eventually we hope to model a number of galaxies in the
GOALS sample of luminous infrared galaxies \citep{A+09}.

\section{IDENTIKIT~2}

IDENTIKIT~2 \citep{B11} uses the self-consistent plus test-particle
simulations described above to \textit{solve} for the initial
orientation of each disk.  This offers a significant shortcut compared
to the trial and error technique used with IDENTIKIT~1.  The main
assumption required is that the tidal features associated with each
galaxy can be traced back to a single disk with a unique orientation.

To see how IDENTIKIT~2 works, consider a single region of phase space
with finite extent in $X$, $Y$, and $V_Z$ and infinite extent in the
remaining dimensions.  This region, hereafter called a ``box'', is
placed so as to sample the tidal material from a particular galaxy.
At some time post-encounter, a single pass through the test-particle
array for that galaxy selects \textit{all} particles falling within
the box.  Each particle has been labeled with its initial spin axis
relative to its parent galaxy, so the selected particles define a
density distribution on the unit sphere of all possible spin
directions.  As a rule, this distribution is extended and does not
define a unique orientation for the parent disk.  However,
\textit{another} box sampling tidal material from the same disk will
generate a different distribution, and the two distributions must
overlap at the disk's true orientation.

Given boxes tracing tidal features from both galaxies, IDENTIKIT~2 can
solve for the inclinations and azimuths of both disks, directly
determining four of the sixteen encounter and viewing parameters.  The
remaining twelve, however, must be specified beforehand, and if the
specified values are wrong then the derived disk orientations will
probably be wrong as well.  But if \textit{at least} three boxes --
ideally more -- are used to sample each disk, the effects of parameter
mismatch tend to destroy the mutual overlap of the distributions.  By
quantifying how well the distributions overlap -- for example, by
forming the product of the densities they trace -- a relative figure
of merit for different solutions is obtained.

With a fairly robust method of measuring quality of fit, it's possible
to implement automatic searching over some subset of the twelve
parameters besides disk orientation.  \citet{B11} described an
implementation designed for systems composed of two disk galaxies
which have not yet merged.  In this version, just six parameters --
the initial orbit ($p$, $\mu$, $e$), time since pericenter $t$,
velocity scale $\mathcal{V}$, and zero-point $v_{0}$ -- must be
specified ahead of time.  In addition to a set of boxes tracing the
tidal features of each galaxy, the algorithm requires coordinates on
the plane of the sky for both galaxy nuclei.  It performs a blind
search over all possible viewing directions ($\theta_X$, $\theta_Y$);
for each direction, the length scale $\mathcal{L}$, rotation about the
line of sight $\theta_Z$, and position zero-point $\mathbf{r}_0$ are
determined by requiring the centers of the model galaxies to match the
observed positions.  Fits to both disks are scored independently as
described above; the viewing direction which maximizes the product of
both scores is selected.  In tests with a small ensemble of simulated
random mergers, the algorithm reconstructed encounter geometry and
viewing direction with median errors of $< 8^\circ$.

\begin{figure}[t!]
  \begin{center}
    \includegraphics[width=4.25in]{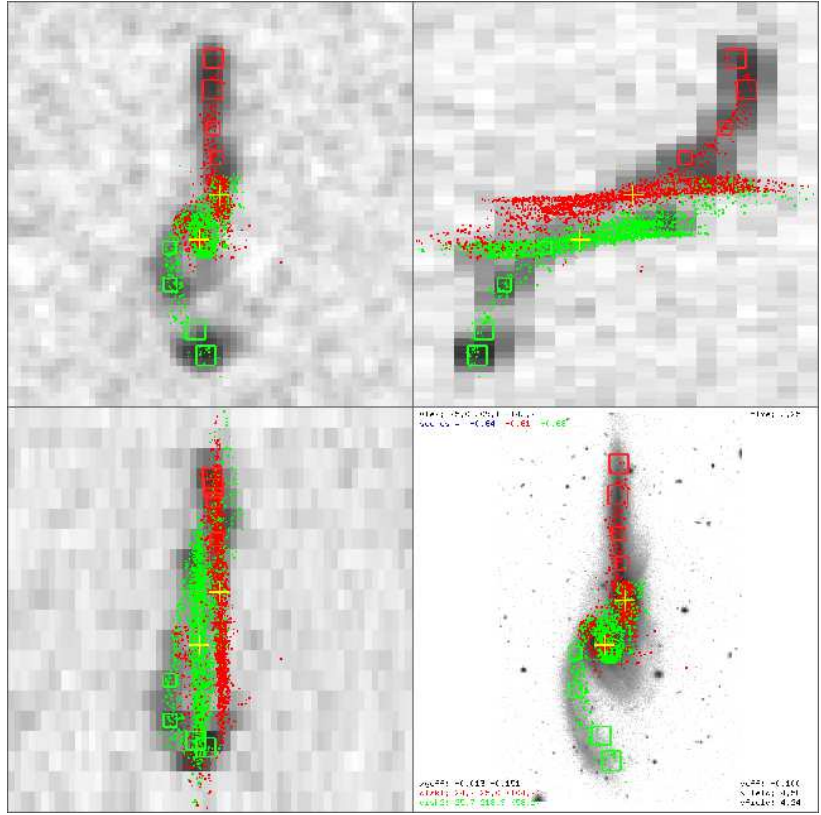}%
  \end{center}
  \caption{NGC~4676.  Top-left and bottom-right panels show H{\small
  I} and optical images, respectively; North is up and West is right.
  Top-right and bottom-left panels show H{\small I} position-velocity
  diagrams; velocity increases to left and down, respectively.  Boxes
  show constraints used to determine disk orientation.  Points show
  IDENTIKIT~2 model.}%
  \label{fig2}%
\end{figure}

Fig.~\ref{fig2} shows that the present algorithm already has
interesting real-world applications.  Here, four regions are allocated
to each tail of ``The Mice'', NGC~4676; these regions track the
H{\small I} morphology and line-of-sight kinematics of the tails
\citep{HvG96}.  In addition, the solution was constrained by requiring
the simulated nuclei (crosses) to match the observed positions, and
fall within the observed range of systemic velocities.  Six parameters
were specified a priori: the orbital eccentricity ($e = 1$), mass
ratio ($\mu = 1$), pericentric separation ($\sim 4.5$ disk scale
lengths), time since pericenter ($\sim 1.25$ disk rotation periods),
velocity scale and velocity zero-point are close to the values adopted
in earlier models \citep[e.g.][]{B04}.  However, the remaining ten
parameters were all derived by the algorithm, which required about two
minutes on a $2.16 \,\mathrm{GHz}$ processor to examine $5120$ viewing
directions and produce the model shown here.  This model is about as
good as models of NGC~4676 derived by hand.  Other ``well-separated''
systems which can be modeled in the same way include Arp~256, Arp~295,
and Arp~298; H{\small I} data is available for all three.

\subsection{Uniqueness and uncertainty}

\begin{figure}[t!]
  \begin{center}
    \includegraphics[angle=-90,width=5.0in]{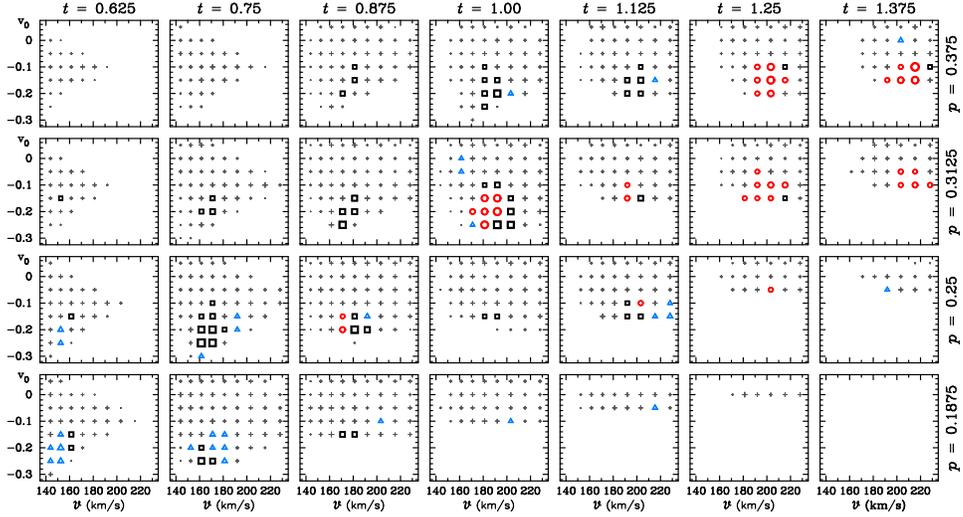}%
  \end{center}
  \caption{Fourteen-parameter search for models of NGC~4676.  From one
  panel to the next, time since pericenter $t$ increases left to
  right, and pericentric separation $p$ increases bottom to top.  Each
  panel shows a $9 \times 8$ grid of models in $\mathcal{V}$ and
  $v_{0}$.  Symbol size indicates model score.  The $122$
  highest-scoring solutions are visually classified as poor
  (triangles: $N = 30$), fair (squares: $N = 56$), and good (circles:
  $N = 36$); low-scoring solutions are shown as grey crosses.}
  \label{fig3}%
\end{figure}

IDENTIKIT~2 is fast and robust enough to go beyond the ten-parameter
search described above.  For a system like NGC~4676, it's reasonable
to fix the mass ratio $\mu = 1$, adopt an $e = 1$ (parabolic) initial
orbit, and perform a blind search in pericentric separation $p$, time
since pericenter $t$, velocity scale $\mathcal{V}$, and velocity
zero-point $v_{0}$, as well as the angles $\theta_X$ and $\theta_Y$
defining direction of view.  As the number of parameters to be
determined increases, the algorithm yields solutions with comparable
overall scores for many parameter combinations.  In Fig.~\ref{fig3}
solutions have been visually graded as poor, fair, or good
representations of NGC~4676.  It's clear that numerical score is an
imperfect indicator of quality, since the highest-scoring solutions
(in the $t = 0.75$, $p = 0.25$ panel) are classified as fair, but many
good solutions do get high scores, while poor solutions consistently
get low scores.

The high-scoring solutions in Fig.~\ref{fig3} are not scattered at
random; most fall along a diagonal from lower left to upper right
across the panels, indicating a general correlation between time since
pericenter $t$ and pericentric separation $p$.  Other parameters which
correlate with $t$ include the velocity zero-point $v_{0}$, viewing
angles ($\theta_X$, $\theta_Y$), and length and velocity scale factors
$\mathcal{L}$ and $\mathcal{V}$.  These correlations indicate that
acceptable solutions populate an ``error ellipse'' in parameter space.
As a group, these solutions are relatively homogeneous; they all
appear to originate from a single connected region of parameter space.
Moreover, the physical time since pericenter, $T =
(\mathcal{L}/\mathcal{V}) t$, is very well constrained; all of the
good solutions and almost all of the fair ones yield $T$ values
between $150$ and $200 \,\mathrm{Myr}$.

Although it may seem better to obtain a unique model, the ensemble of
solutions for NGC~4676 shown in Fig.~\ref{fig3} is probably a good
representation of the actual uncertainties inherent in modeling this
system with available H{\small I} data.  There's no telling if most
interacting galaxies will likewise yield fairly well-constrained
solutions; NGC~4676 is a relatively simple system, and others may be
harder to constrain.

This 14-parameter search shows that comprehensive surveys of the
encounter parameter space are possible.  While the \textit{internal}
structures of the victim galaxies remain to be parametrized (\S~4),
the ability to search such large spaces is encouraging.

\section{Mass models}

The results presented above use a generic galaxy model with a
bulge+disk:halo mass ratio $(m_\mathrm{b} + m_\mathrm{d}) \!:\!
m_\mathrm{h} = 1\!:\!4$.  While this model is adequate for some
purposes, the halo comprises just $80$\% of the total, which is less
than expected in CDM cosmologies.  Moreover, any effort at data-driven
modeling of real mergers must address variations in initial galaxy
structure.  How sensitive are model results to inevitable
discrepancies between the adopted and actual structure of the
progenitor galaxies?  Can tidal encounter models accurately probe the
overall depth and structure of halo potential wells?

\begin{figure}[t!]
  \begin{center}
    \includegraphics[angle=-90,width=5.0in]{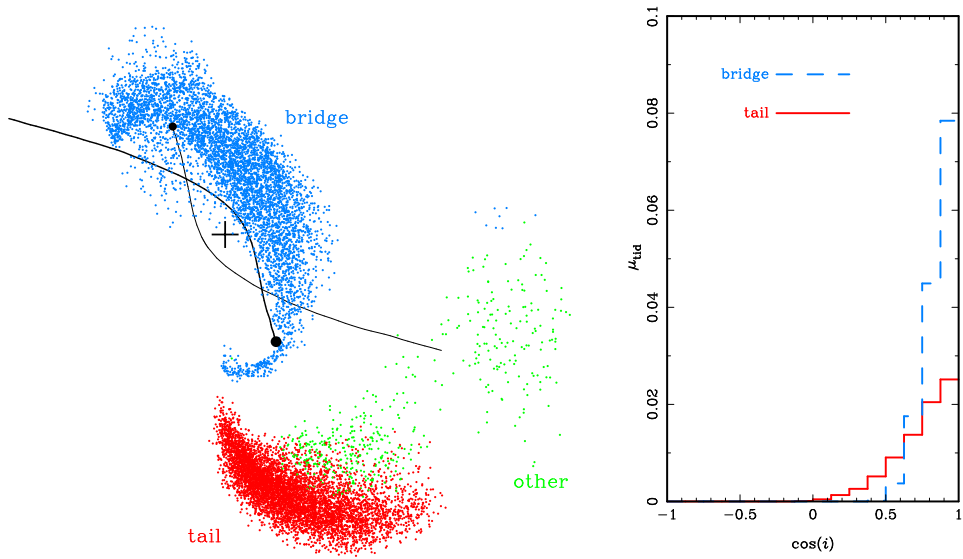}%
  \end{center}
  \caption{Left: result of a moderately close passage between two
  equal-mass galaxies, seen shortly after apocenter.  Solid lines are
  galaxy trajectories; the cross marks the system's center of mass.
  Points show tidal structures originating from the galaxy marked by
  the larger filled circle, integrated over all possible disk
  orientations; bridge and tail particles populate distinct regions as
  shown.  Right: histogram showing bridge and tail mass fractions as
  functions of disk inclination $i$.}
  \label{fig4}%
\end{figure}

Existing work on the effects of halo structure on tidal features
\citep{DMH96, DMH99, SW99} has largely focused on direct, co-planar
encounters which maximize tidal features; a systematic study varying
encounter geometry as well as galaxy structure is bedeviled by a large
number of parameters.  However, IDENTIKIT offers an efficient way to
treat encounter geometry, since a single simulation simultaneously
models all possible disk orientations.  Fig.~\ref{fig4} presents an
example, based on a galaxy model with mass ratio $(m_\mathrm{b} +
m_\mathrm{d}) \!:\!  m_\mathrm{h} = 1\!:\!9$.  The plot shows test
particles which have attained a maximum distance from their parent
galaxy of at least three times their initial orbital radius; these
particles populate tidal features.  Particles are classified as
belonging to bridge or tail structures based on their position
relative to the two galaxies at (1) first passage and (2) instant of
maximum distance; a very few particles don't belong to either.  No
selection for initial disk orientation is done, so the result is a
``synoptic'' view of \textit{all} tidal structures resulting from this
encounter.

The histogram on the right of Fig.~\ref{fig4} shows how the
mass fractions $\mu_\mathrm{tid}$ in bridges and tails depend on disk
inclination $i$.  At small inclinations ($i < 30^\circ$), the bridge
is relatively massive, comprising $\sim 8$\% of the disk material.
However, bridge mass drops rapidly with increasing $i$.  The tail is
less massive, amounting to $\sim 2.5$\% of the disk material for small
$i$, but drops off less rapidly with increasing $i$.  In effect,
bridge-to-tail mass ratio is a decreasing function of inclination;
this trend would be very difficult to quantify using conventional
simulations.

\bibliography{barnes_arXiv}

\end{document}